\documentclass[aps,pre,preprint,showpacs]{revtex4-1}
\usepackage{ifpdf}
\usepackage{pdfsync}
\ifpdf
\usepackage{hyperref}
\else
\usepackage[hypertex]{hyperref}
\fi
\usepackage{graphicx,subfigure}
\usepackage{amsmath,amsfonts,amssymb}

\newcommand{\f}{\frac}

\newcommand{\intl}{\int\limits}

\graphicspath{{D:/localtexmf/Mytex/LGH/mathplots/},{D:/localtexmf/Mytex/LGH/eqstcan/zeno_jcp/zeno_timmerman/}}
\begin{document}
\title{The critical compressibility factor of fluids from the Global Isomorphism approach}
\author{V.~L. Kulinskii}
\email{kulinskij@onu.edu.ua}
\affiliation{Department for Theoretical
Physics, Odessa National University, Dvoryanskaya 2, 65026 Odessa, Ukraine}
\begin{abstract}
The relation between the critical compressibility factors $Z_{c}$ of the Lennard-Jones fluid and the Lattice Gas (Ising model) is derived within the global isomorphism approach. On this basis we obtain the alternative form for the value of the critical compressibility factor which is different from widely used phenomenological Timmermans relation. The estimates for the critical pressure $P_c$ and $Z_c$ of the Lennard-Jones fluid are obtained in case of two and three dimensions. The extension of the formalism is proposed to include the Pitzer's acentric factor into consideration.
\end{abstract}
\pacs{64.60.Bd, 05.50.+q, 05.20.Jj} \maketitle
\section{Introduction}\label{sec_intro}

There are a number of phenomenological facts known for liquids which can be formulated in simple terms but their proof or explanation from the first principles of Statistical Theory of Liquid State is absent. The remarkable
linearities of the binodal diameter \cite{crit_diam0} and the unit compressibility curve $Z=1$ (Zeno-line) \cite{eos_zenobenamotz_isrchemphysj1990}
in density-temperature plane are probably of the oldest known in liquid state physics.

The law of rectilinear diameter (LRD) of the binodal is as following:
\begin{equation}\label{rdl}
  n_{d} = \f{n_{l}+n_{g}}{2\,n_c} = 1+A\,
\left(\, 1-T/T_c \,\right)\,,\quad T<T_c\,,
\end{equation}
where $n_{l,g}$ are the number densities of the liquid and vapor phases correspondingly, $n_{c}, T_c$ are the critical density and temperature. The LRD is observed for simple molecular fluids in wide temperature region of liquid-vapor coexistence. It is also confirmed by the computer simulations (see e.g. \cite{crit_liqvamiepotent_jcp2000}) and commonly serves as a tool to estimate the critical density \cite{book_frenkelsimul}.

Another remarkable phenomenological linearity established early before for the van der Waals equation of state by Batschinsky \cite{eos_zenobatschinski_annphys1906} and confirmed for many fluids \cite{eos_zenoholleran_jcp1967}. The fact is that the states with $Z=P/(n\,T) = 1$ form the straight line on $n-T$ plane:
\begin{equation}\label{z1}
Z = \f{P}{n\,T} =1 \quad \Rightarrow \quad \f{T}{T_{Z}}+\f{n}{n_Z} = 1\,.
\end{equation}
This line is called the Zeno-line \cite{eos_zenobenamotz_isrchemphysj1990}.
The parameters $T_{Z}$ and $n_{Z}$ are determined as follows:
\begin{equation}\label{tbnb}
  B_2(T_{Z}) = 0\,,\quad n_{Z}= \f{ T_{Z} }{B_3\left(\,T_{Z}\,\right)}\,\left. \f{dB_2}{dT}\right|_{T= T_{Z}}\,.
\end{equation}
As the available data show  the linear law \eqref{z1} holds for real fluids and model systems with pair interactions of the Lennard-Jones (LJ) type with good accuracy. Moreover, it holds even for associative liquids like ammonia and water \cite{eos_zeno_jphyschem1992,eos_zenoboyle_jphysc1983,*water_zenoline_ijthermophys2001}.
The main difficulty in the explanation of these linearities is the construction of adequate equation of state (EoS) for real substances. In rigorous sense it is unsolvable problem since the EoS is complicated functional of the interparticle interactions. Nevertheless, due to principle of corresponding states (PCS) it is possible to introduce the classes of the thermodynamic similarity \cite{pcs_guggenheim_jcp1945,book_filippovthermophys_1988en}. In its general statement the PCS is rather trivial since it can be proven explicitly only if the interaction potentials are of the same functional form modulo simple scaling of the parameters \cite{book_prigozhisolut}. For real systems the refinement of the na\"{\i}ve form of the PCS is needed and additional invariants are introduced \cite{book_filippovthermophys_1988en,eos_riedelpcs1_cingtech1954,*liq_pcspitzer2_jamchemsoc1955}. The critical compressibility factor (CCF) $Z_{c}  = P_{c}/(n_c\,T_c)$ is the simplest scaling invariant widely used in the formulation of the PCS. For the vdW EoS $Z_c = 3/8$. Simple fluids, e.g. nobel gases $Ar$, $Kr$, $Xe$, have $Z_{c}\approx 0.28 \div 0.3$. For the CCF the empirical Timmerman's relation is known:
\begin{equation}\label{ztimm}
  Z_{c}\approx \f{n_{c}}{n_{Z}}\,\,.
\end{equation}
It holds quite good for many substances \cite{book_filippovthermophys_1988en,eos_zenoapfelvorob_jpcb2013}. To the best of our knowledge no theoretical explanations for this relation exist up to now.

In this paper the relation between the critical compressibility factors of the LJ-fluid and the lattice gas is obtained. We use the global isomorphism approach proposed in \cite{eos_zenomeglobal_jcp2010} and the results of \cite{eos_zenomeunified_jphyschemb2011}.
We also obtain the value of $Z_c$ for the LJ-fluid in 2D and  3D cases. The result for $Z_c$ derived below provides the alternative to the pure \textit{empirical} Timmermans relation \eqref{ztimm}. It grounds on the equality of thermodynamic potentials of LJ-fluid and lattice gas \cite{eos_zenomeunified_jphyschemb2011}. The structure of the paper is as follows. In Section II we briefly outline the global isomorphism approach between the LJ-fluid and the lattice gas (Ising model). Section III is devoted to the derivation of the relation between critical compressibility factors $Z_c$ of LJ-fluid and lattice gas. On this basis we get the estimates for the critical pressures of LJ-fluid in two and three dimensions. In Section IV we describe simple way of how to include the Pitzer's acentric factor into consideration. The results are summarized and discussed in conclusion.

\section{Global isomorphism relations}
The extension of the PCS, which unifies the
equilibrium thermodynamic properties of the liquid metals and
molecular fluids, was put forward in series of works of Apfelbaum
and Vorob'ev \cite{eos_zenoapfelbaum_jpchemb2006,*eos_zenoapfelbaum1_jpcb2009,*eos_zenoline_potentials_jcp2009}. The authors introduced the concept of the triangle of liquid-gas
states.

As was shown in \cite{eos_zenome_jphyschemb2010} the results of
\cite{eos_zenoapfelbaum_jpchemb2006,*eos_zenoapfelbaum1_jpcb2009,*eos_zenoline_potentials_jcp2009}
can be cast into the compact form of projective mapping between
the density-temperature phase diagrams of the fluid and the
lattice gas:
\begin{equation}\label{projtransfr_my}
n =\,n_*\,\f{x}{1+z \,\tilde{t}}\,,\quad  T =\, T_*\,\f{z\,
\tilde{t}}{1+z \,\tilde{t}}\,,\quad z = T_{c}/(T_{*}-T_{c})\,,
\end{equation}
where $x$ and $t$ are the density and the temperature of the lattice gas (LG), $\tilde{t}=t/t_c$, $n_*$ and $T_*$ are some parameters which will be defined below (see also \cite{crit_globalisome_jcp2010}). The coordinates of the CP of the lattice gas are normalized so that $\tilde{t}_c = 1$ and $x_c = 1/2$. The derivation of \eqref{projtransfr_my} uses simple geometrical arguments and substantially relies on the LRD and Zeno-linearity\cite{eos_zenome_jphyschemb2010,eos_zenomeunified_jphyschemb2011}.
Eq.~\eqref{projtransfr_my} maps the states within the liquid-gas triangle on the $(x,t)$-states of the lattice gas (Ising model). This brings the concept of the global isomorphism between the systems once the LRD and the Zeno-linearity hold \cite{eos_zenomeunified_jphyschemb2011}.

Recently, several generalizations of transformation \eqref{projtransfr_my} have been proposed \cite{eos_zeno_lattice2real_jpcb2010,eos_zenoherschbachktrans_jpcb2013}. We will not use these extensions here and prefer to use Eq.~\eqref{projtransfr_my} as the ``minimal`` and the simplest  form of the isomorphism transformation compatible with the LRD and the Zeno-linearity based on the results of previous works \cite{eos_zenomeglobal_jcp2010,crit_globalisome_jcp2010}.

In global isomorphism approach the notion of Zeno-element is used instead of the Zeno-line \cite{eos_zenomeglobal_jcp2010}. The Zeno-element is the line of the same form as Eq.~\eqref{z1} with the parameters $T_{Z}\to T_{*}\,, n_{Z}\to n_{*}$ determined via the Boyle point in van der Waals (vdW) approximation:
\begin{equation}\label{tbvdwmy}
  B^{vdW}_2(T_{*}) = 0\,,\quad  T_{*} =  T^{(vdW)}_B  = \f{a}{b}\,,
\end{equation}
and
\begin{equation}\label{nbvdwmy}
n_*= \f{ T_* }{B_3\left(\,T_*\,\right)}\,\left. \f{dB_2}{dT}\right|_{T= T_*}\,.
\end{equation}
Here
\begin{align}\label{vdw_ab}
a =\,\, -2\pi\,\intl_{\sigma}^{+\infty}\Phi_{attr}(r)\,r^2\,dr
\end{align}
and $\Phi_{attr}(r)$ is the attractive part of the full
potential $\Phi(r)$, $\sigma$ is the effective diameter of the particle so that $b = \f{2\pi}{3}\,\sigma^{3}$. Note that we do not use the van der Waals EoS, rather we use the van der Waals approximation for the Boyle temperature \cite{book_ll5_en}. Therefore only attractive part of the interaction potential is used with the hard sphere radius cut-off. From the physical point of view it correlates with the fact that the lattice gas interaction includes the attraction only. The repulsive part is introduced in the model only via the restriction rule ``one site - one particle``. In \cite{crit_globalisome_jcp2010} this approach was applied for two and three dimensional LJ fluid. The parameters $T_*,n_*$ of the mapping show good agreement with theoretical values \eqref{tbvdwmy}, \eqref{nbvdwmy} under fitting the available data. The parameter $z$ of the transformation can be related with the exponent of the attractive part of the potential $\Phi_{att}(r) \sim r^{-n}$: $z = d/n$, where $d$ is the dimension \cite{eos_zenomeglobal_jcp2010}.
Also Eq.~\eqref{projtransfr_my} gives pretty good estimates for the locus of the CP of the Lennard-Jones fluids \cite{eos_zenomeglobal_jcp2010}.

The authors of \cite{eos_zenoherschbachktrans_jpcb2013} have shown that the parameters $T_{c}/T_Z, \, n_{c}/n_{Z}$ can be related with the Pitzer's acentric factor $\omega$ which accounts for the molecular shape. This puts the question of taking into account the hard core effects of the repulsive part of the interaction in the parameter $z$ of the transformation \eqref{projtransfr_my}. This opportunity needs further studies and will be discussed briefly in Section IV.

\section{The compressibility factor of the fluid and the alternative of the Timmermans relation}
Here we consider the critical compressibility factor $Z_c$, which  is the core parameter for the formulation of the PCS, and get the analog of the Timmermans relation in the global isomorphism context. To do this we derive the relation between thermodynamic potentials of the fluid $J = P(\mu,T)\,V$ and lattice gas $\mathfrak{G}(h,t) =\mathcal{N}\,g(h,t) = \mathfrak{P}(h,t)\,V$, where $\mathfrak{P}$ is the pressure of the lattice gas. Let us define the number of lattice sites $\mathcal{N}$ with the volume of the fluid by $\mathcal{N} = n_*\,V$. We start with the basic thermodynamic relations:
\begin{equation}\label{eq_densities}
  n = \left.\frac{\partial P}{\partial \mu}\right|_{T}\,,\quad x = \left.  \frac{\partial g}{\partial h}\right|_{t}
\end{equation}
and from Eq.~\eqref{projtransfr_my} we get:
\begin{equation}\label{pderiv}
\left.\frac{\partial P}{\partial \mu}\right|_{T} =  \f{n_*}{1+z\,t}\,  \left.  \frac{\partial g}{\partial h}\right|_{t}\,\,.
\end{equation}
As the global isomorphism relates  the thermodynamic states of corresponding systems then the conjugated fields $h$ and $\mu$ can be related $\mu = \mu(h,t)$ due to simple relation:
\begin{equation}\label{muderiv}
\left.\frac{\partial }{\partial \mu}\right|_{T} =
\left.  \frac{\partial h}{\partial \mu}\right|_{T}\,\left.\frac{\partial }{\partial h}\right|_{t}
\end{equation}
Note, that here is important that the temperature variable of the fluid $T$ depends only on the temperature of the lattice gas $t$.
Comparing Eq.~\eqref{muderiv} with Eq.~\eqref{pderiv} we get the relation:
\begin{equation}\label{muh}
  \mu -\mu_0(T)  = h\,\left(\,1+z\,\tilde{t} \,\right)\,\,.
\end{equation}
Here $\mu_0(T)$ is the value of the chemical potential on the saturation curve below the critical point $T<T_c$.
The density is nothing but the derivative of the thermodynamic potential on the conjugated field. This fact together with the linear character of \eqref{projtransfr_my} on densities $n$ and $x$ leads to the relation between pressures of fluid $P$ and of the LG $\mathfrak{P}$:
\begin{equation}\label{pp}
  P(T,\mu) = \mathfrak{P}(t(T),h(\mu,T))\,,
\end{equation}
or in terms of the potentials:
\begin{equation}\label{potpot}
  P(T,\mu)\,V = \mathfrak{G}(t(T),h(\mu,T))\,,
\end{equation}
It is easy to check that Eq.~\eqref{projtransfr_my}follows from Eq.~\eqref{pp} and Eq.~\eqref{muh}.

With this Eq.~\eqref{pp} allows to derive the relation between the critical compressibility factors $Z_c$ of a fluid and the corresponding lattice model. Indeed, from Eq.~\eqref{projtransfr_my} and Eq.~\eqref{pp} we get:
\begin{equation}\label{zmy}
  Z^{(fl)}_{c} = \f{P_c}{n_c\,T_{c}} =\f{(1+z)^2}{z}\,\f{t_c}{T_*}\, Z^{(LG)}_{c}\,\,,
\end{equation}
where $t_c$ is the critical temperature of the Ising model and $Z^{(LG)}_{c}= 2\mathfrak{P}_c/(n_*\,t_c)$ is the CCF of the LG. In accordance with \cite{book_rice_thermodyn} it is related with the partition function per spin $G^{1/N}$ of the Ising model
$Z^{(LG)}_{c} = 2\,\ln G_{c}^{1/N}$.
The value $G_{c}^{1/N}$ can be obtained for lattice models using (high)low-temperature expansions
\cite{crit_dombsykes_prc1956,crit_montroll1968lectures}. E.g. in 3D case $t_c \approx 4.51\,J$ \cite{crit_3disingliufisher_physa1989} and $Z^{(LG)}_{c} = 0.246$ \cite{book_rice_thermodyn}.

Note that Eq.~\eqref{ztimm} and Eq.~\eqref{zmy} differ essentially. In contrast to the Timmermans relation its global isomorphism analog \eqref{zmy} appeals directly to the characteristics of the critical state of the Ising model. This reflects the fact of the isomorphism of critical behavior of the fluid the Ising model \cite{book_stanley}. Moreover, it relates \emph{nonuniversal} parameters of the critical state of these systems (see also \cite{eos_zenomeunified_jphyschemb2011}) which isomorphic in a sense of fluctuational theory of critical phenomena \cite{book_patpokr}.
According to Eq.~\eqref{zmy} the CCF depends on the temperature scale $T_*$ but not on $n_{*}$. To consider the case of the LJ-fluid we take into account that for the LJ potential:
\begin{equation}\label{lj}
  V_{LJ}(r) = 4\,\varepsilon\,
\left(\, \left(\, \f{\sigma}{r}  \,\right)^{12} - \left(\, \f{\sigma}{r}  \,\right)^{6}\,\right)\,\,,
\end{equation}
$T_{*} = 4\,\varepsilon$ and the relation between the LJ parameter $\varepsilon$ and the spin-spin interaction $J$ of the Ising model is $\varepsilon = 4\,J$ \cite{book_baxterexact}.
We assume that the LJ parameter $\varepsilon$ can be identified with the interaction constant of the lattice gas. This is based on the fact that $\varepsilon$ is the energy of the LJ interaction at equilibrium distance. The latter can be identified with the period of the cubic lattice.  Further we use conventional dimensionless units for the temperature $T\to T/\varepsilon$ density $n\to n\sigma^3$ and pressure $P \to P\,\sigma^3/\varepsilon$.

Calculation of the factors in \eqref{zmy} results in:
\begin{equation}\label{zmy_lj3d}
  Z^{(fl)}_{c}\approx 1.27\, Z^{(LG)}_{c} = 0.281\,,
\end{equation}
which correlates well with the CCF value for
simple fluids. The corresponding estimate for the value of critical pressure is:
\begin{equation}\label{pc_my}
 P_c = 0.121\,,
\end{equation}
and shows good correspondence with the result of simulations \cite{eos_lj_molphys1992} $P_c = 0.126$ for the 3D LJ fluid.

Our main result \eqref{zmy} can be also applied in 2D case. Due to the Onzager's solution we know the basic quantities exactly:
\[Z^{(LG)}_{c} \approx 0.097\,,\quad t_c \approx 2.27\,J\,,\]
we get the estimate:
\begin{equation}\label{zmy_lj2d}
  Z^{(fl)}_{c}\approx 1.513\, Z^{(LG)}_{c} = 0.146\,.
\end{equation}
The value of the critical pressure of the 2D LJ fluid is:
\begin{equation}\label{pcmy_lj2d}
P_{c} =   Z^{(fl)}_{c}\,n_{c}\,T_{c}\approx 0.026\,,\quad n_{c} = 0.353\,,\quad T_{c} = 0.5\,.
\end{equation}
The estimates for critical density and temperature are consistent with known results for two dimensional LJ fluid \cite{crit_lj2dim_jcp1991}. The values of $Z_c$ and $P_c$ lower than those obtained within perturbation theory $P_c=0.046$ and in simulations $P_c = 0.037$ \cite{eos_2dljabraham_physrep1981}. Possible explanation of this fact is that these approaches underestimate the flatness of the binodal of the 2D LJ fluid. This is  due to the global power law behavior with the critical exponent $\beta = 1/8$ because of the global isomorphism with the 2D Ising model \cite{crit_globalisome_jcp2010}. For convenience all the parameters and the results obtained are summarized in Table~\ref{tab1}.
\begin{table}
  \centering
  \begin{tabular}{|c|c|c|c|c|c|c|c|}
    \hline
LJ ``6-12`` fluid &
\hspace{0.2cm} $T_*$ \hspace{0.2cm}&
\hspace{0.2cm} $n_*$ \hspace{0.2cm}&
\hspace{0.2cm} $T_c$ \hspace{0.2cm}&
\hspace{0.2cm} $n_c$ \hspace{0.2cm}&
\hspace{0.2cm} $P_c$ \hspace{0.2cm}&
\hspace{0.2cm} $Z^{(fl)}_c$ \hspace{0.2cm}&
\hspace{0.2cm} $Z^{(LG)}_c$ \hspace{0.2cm}\\
\hline
$2D$ &2&0.941&0.5&0.353&0.026&0.147&0.097\\
\hline
$3D$ &4&0.976&1.333&0.325&0.122&0.281&0.246\\
\hline
  \end{tabular}
  \caption{The results for the critical point parameters of $LJ$ fluid in 2D and 3D cases. For 2D case $z = 1/3$ and $z=1/2$ in 3D \cite{eos_zenomeglobal_jcp2010}.}\label{tab1}
\end{table}

\section{Simple modification of the approach with the acentric factor}
\begin{figure}
\includegraphics[scale=0.5]{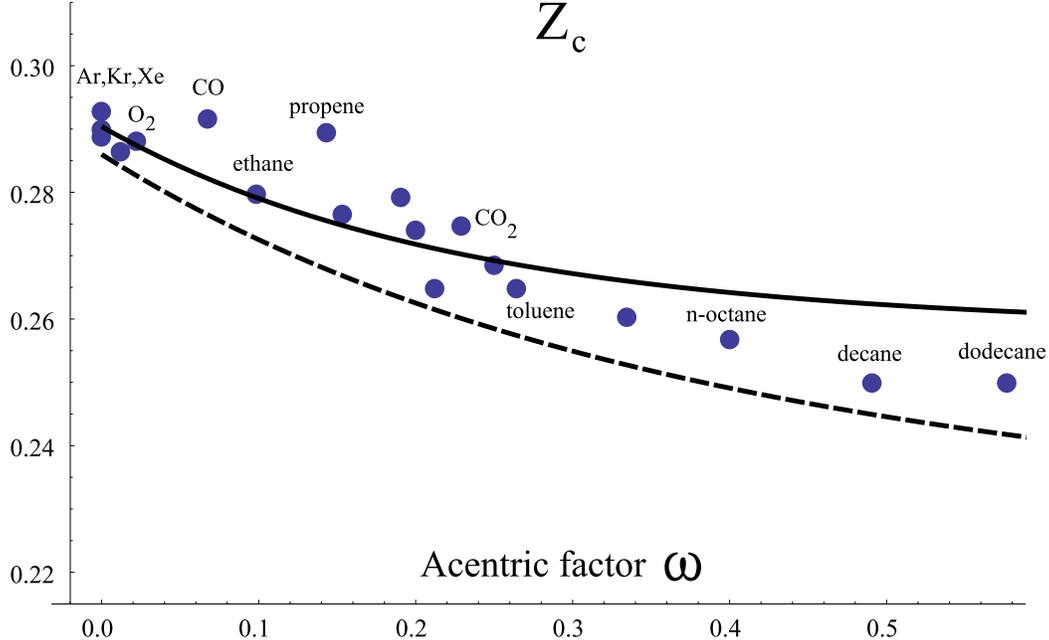}\\
  \caption{The dependence $Z_c$ given by Eq.~\eqref{zmy_omega}
(see text) on the acentric factor $\omega$ (solid line). For the comparison the result of Eq.~\eqref{zc_hercsh} (see \cite{eos_zenoherschbachktrans_jpcb2013}) is shown by dashed line.}\label{fig_zcmy}
\end{figure}

Naturally, the modification of the global isomorphism approach is needed in case of more complex fluids. From such point of view the thermodynamic class of similarity is determined by both a) the structure of the lattice of the isomorphic model and b) the correction of parameter $z$ due to repulsive part of the interaction. As to the first point the existence of such lattice model is provided by the LRD and the Zeno-linearity. The second point rises the problem to include  into consideration other invariants like Pitzer's acentric factor $\omega$ or Riedel parameter \cite{eos_zenoherschbachktrans_jpcb2013}.
Such parameters on the phenomenological level represent the shape of the molecules. Therefore they can be related with the hard-core effects due to repulsive part of the potential. This issue needs further studies. Here we briefly outline the simplest phenomenological way. The idea is to use Eq.~\eqref{zmy} with  the parameter $z$ dependent on the acentric factor $\omega$ so that:
\begin{equation}\label{zmy_omega}
  Z^{(fl)}_{c}(\omega) = \f{P_c}{n_c\,T_{c}} =\f{(1+z(\omega))^2}{z(\omega)}\,\f{t_c}{T_*}\, Z^{(LG)}_{c}\,\,,
\end{equation}
where
\begin{equation}\label{zw0}
  z(\omega) = \f{1}{T_*/T_c(\omega)-1}\,\,.
\end{equation}
To introduce $\omega$-dependence for Eq.~\eqref{zw0} in our approach we use the result of \cite{eos_zenoherschbachktrans_jpcb2013} for the parameter $T_{Z}/T_c$:
\begin{equation}\label{tw}
T_{Z}/T_c(\omega) = 2.1 + 0.63\,\exp(-3.8\,\omega)\,\,.
\end{equation}
As has been noted above $T_{*}\ne T_{Z}$. So we introduce $a=T_{*}/T_{Z}$ as the fitting parameter:
\begin{equation}\label{zw1}
  z(\omega,a) = \f{1}{a\,T_Z/T_c(\omega)-1}\,\,.
\end{equation}
where $T_Z/T_c(\omega)$ is given by Eq.~\eqref{tw}. The result of the best fit to the data gives $a \approx 1.187$ and shown in Fig~\ref{fig_zcmy} by solid line.
If one takes the value $T_Z = 3.416$ for the Boyle temperature of the LJ-fluid then $T_* \approx 4.06 $, which is in perfect agreement with the value of the Boyle temperature in vdW approximation $T_* = 4$. This shows self-consistency of the procedure based on representation Eq.~\eqref{zw1} with Eq.~\eqref{projtransfr_my} at least for the substance with low ($\omega <0.3 $) acentric factor. For comparison in  Fig.~\ref{fig_zcmy} we represent the data for the values of $Z_c$ and corresponding acentric factor $\omega$ \footnote{We would like to thank unknown referee who kindly sent the plot of $Z_c(\omega)$ based on the results of \cite{eos_zenoherschbachktrans_jpcb2013}}. Here the points are the values of $Z_c$ according to the CP locus data (see \cite{eos_zenoherschbachktrans_jpcb2013}). Dashed curve is the fitting curve of the form
\begin{equation}\label{zc_hercsh}
  Z_c(\omega) \approx \f{n_{c}}{n_Z} = \f{1}{4.41 - 0.91\,\exp
\left( -2.1\,\omega \right)}\,,
\end{equation}
which is based on the Timmermanns relation \eqref{ztimm} and the results of \cite{eos_zenoherschbachktrans_jpcb2013}.

For the substance with higher values of $\omega$ and low values of $Z_c$ further improvement should be related with the lattice structure of the isomorphic lattice model. Obviously, this is because simple cubic lattice is oversimplified model for the fluids with molecules of highly nonspherical shape.

\section*{Conclusion}
In conclusion we discuss the main result of the work and comment on further developments. We have derived the relation \eqref{zmy} between the CCF of the LJ fluid and that of the lattice model of Ising like type. The obtained relation reproduces the known value of $Z_c$ for simple liquids basing on the available data for the 3D Ising model. The analogous relation can be derived for more complex fluids. In view of the global isomorphism approach they are those fluids for which the LRD and the Zeno linearity holds with good accuracy. According to the results of \cite{eos_zenoapfelbaum_jpchemb2006,*eos_zenoapfelbaum1_jpcb2009,*eos_zenoline_potentials_jcp2009} water and liquid alkali metals can be considered this way though the CCF is much lower ($Z_{c}= 0.19\div 0.23$) than the value $Z_c$ of simple fluids. According to simple PCS these liquids belong to other class of the thermodynamic similarity. Yet the equilibrium thermostatic properties of these fluids are mainly formed by the polarizational short range forces. Thus proper generalization of the PCS, e.g. in the framework of the global isomorphism approach can be elaborated e.g. in order to take into account the associative features of such fluids.

\section*{Acknowledgement}
The author thank Konstantin Yun for support during the work on the manuscript and Prof. V.~S. Vorob'ev for kindly sending some recent papers relevant to the subject. The unknown referee is acknowledged for providing the results for $Z_c(\omega)$ based on the data of \cite{eos_zenoherschbachktrans_jpcb2013} and stimulating discussion as to the inclusion of acentric factor $\omega$ into consideration.

\newpage
%

\end{document}